\documentclass[aps,prd,twocolumn,showpacs,superscriptaddress,groupedaddress,eqsecnum,nofootinbib,floatfix]{revtex4-2}

\usepackage{amsmath}
\usepackage{amsfonts}
\usepackage{amssymb}	
\usepackage{graphicx}
\usepackage{bm}
\usepackage{color}
\usepackage{commath}

\usepackage{graphicx}
\usepackage{dcolumn}

\usepackage{color}
\usepackage{xcolor}
\definecolor{linkcolor}{rgb}{0.0,0.3,0.5}
\usepackage[unicode,colorlinks=true,citecolor=linkcolor,linkcolor=linkcolor,urlcolor=linkcolor]{hyperref}

\usepackage{cleveref}

\graphicspath{{../figures/}}

\begin{document}

\title{Nonlocal-in-time effective one body Hamiltonian in scalar-tensor gravity at third post-Newtonian order}

\author{Tamanna \surname{Jain}}
\email{tj317@cam.ac.uk}
\affiliation{Department of Applied Mathematics and Theoretical Physics,University of Cambridge,Wilberforce Road CB3 0WA Cambridge, United Kingdom.}%
\date{\today}
\begin{abstract}
	
We complete the nonlocal-in-time effective-one-body (EOB) formalism of conservative dynamics for massless Scalar-Tensor (ST) theories 
at third post-Newtonian (PN) order.  
The nonlocal-in-time EOB Hamiltonian is obtained by mapping the order-reduced Hamiltonian corresponding to the nonlocal-in-time Lagrangian derived in 
[Phys. Rev. D 99, 044047 (2019)]. 
To transcribe the dynamics within EOB formalism, we use a strategy of 
order-reduction of nonlocal dynamics to local ordinary
action-angle Hamiltonian.  We then map this onto the EOB Hamiltonian to determine the nonlocal-in-time ST corrections to the EOB potentials $(A,B,Q_e)$ at 3PN order.
\end{abstract}

\maketitle
\section{Introduction}
In 2015, the direct detection of gravitational waves (GW) by the LIGO-Virgo Collaboration~\cite{Abbott:2016blz} emitted by inspiralling compact binary, opened new avenues for probing the dynamics in strong gravity regime~\cite{Arun:2006yw,Mishra:2010tp,Li:2011cg,Agathos:2013upa,Cornish:2011ys,Berti:2015itd,Yunes:2016jcc}. It is expected that future GW detectors, like the Einstein Telescope~\cite{Maggiore:2019uih} and Cosmic Explorer~\cite{Evans:2021gyd} will shed more light on alternative theories of gravity by constraining the parameters of such theories.

The simplest theory amongst the alternative theories of gravity is the addition of a massless scalar field to GR, scalar-tensor (ST) theories, which are 
are extensively studied~\cite{Damour:1992we,Damour:1993hw,Damour:1995kt,Freire:2012mg,Khalil:2022sii,Gautam:2022cpb}.
The motivation for ST theories is to explain both the accelerated expansion of the universe as $f(R)$-theories \cite{DeFelice:2010aj} as well as UV complete alternate theories of GR. The two-body PN formalism for ST theories has been extensively studied~\cite{Lang:2013fna,Lang:2014osa,Bernard:2018hta,Bernard:2018ivi,Bernard:2019yfz,Schon:2021pcv,Brax:2021qqo,Sennett:2016klh,Bernard:2022noq}.

%
%
%
%

%
The important violations for ST theories arise through the non-perturbative strong field effects in neutron-stars such as spontaneous scalarisation \cite{Damour:1995kt}. Although the current constraints come from the binary pulsar observations, the future GW detections can better constraint the parameters using strong-field information and additional terms in radiation,\ i.e.  dipolar radiation which is not present in GR, due to the scalar extension of GR~\cite{Palenzuela:2013hsa,Sennett:2016rwa,Khalil:2022sii}. 
%

The EOB formalism was introduced to construct analytical waveform templates for GR~\cite{Buonanno:1998gg,Buonanno:2000ef,Damour:2000we,Damour:2008qf,Damour:2014jta,Damour:2015isa,Damour:2016abl}. Recently, the two-body PN dynamics has also been mapped within the EOB formalism to construct waveform templates for ST theories~\cite{Julie:2017ucp,Julie:2017pkb, Jain:2022nxs}. In our previous work~\cite{Jain:2022nxs}, we determined the EOB potentials for the local part of dynamics at 3PN order.
The aim of this paper is to determine the complete nonlocal-in-time EOB potentials following our results of local part in Ref. ~\cite{Jain:2022nxs}
starting from the 3PN nonlocal-in-time Lagrangian of Ref.~\cite{Bernard:2018hta,Bernard:2018ivi}. Hereafter, the companion paper \cite{Jain:2022nxs} will be referred as Paper I.

The paper is organised as follows. In Sec.~\ref{Sec-reminder}, 
we give a summary of results obtained in Paper I. 
Then, in Sec.~\ref{Sec-OrdinaryHamiltonian} we derive the conserved energy for nonlocal-in-time part using two methods, (i) non-order-reduced nonlocal Hamiltonian using nonlocal phase shift, and (ii) order-reduction of nonlocal dynamics to local ordinary action-angle Hamiltonian. Finally, in Sec.~\ref{Sec-EOB} we map the nonlocal-in-time ordinary Hamiltonian 
into an EOB Hamiltonian at 3PN order.

\section{Summary of Previous results}
\label{Sec-reminder}
We consider mono-scalar massless ST theories described by the following action in the Einstein Frame (the scalar field minimally couples to the metric),
\begin{align}
S&=\frac{c^4}{16 \pi G}\int d^4x \sqrt{-g}(R-2g^{\mu \nu} \partial_\mu \varphi \partial_\nu \varphi)\nonumber\\
  &\qquad\qquad\qquad\qquad+S_m[\Psi, {\mathcal{A}(\varphi)}^2g_{\mu\nu}]~,
\end{align}
where $g_{\mu\nu}$ is the Einstein metric, $R$ is the Ricci scalar, $\varphi$ is the scalar field, $\Psi$ collectively denotes the matter fields, $g \equiv \det(g_{\mu\nu})$ and $G$ is the bare Newton's constant~\cite{Jain:2022nxs}.
As Paper I (see, Table I), we adopt the conventions and notations of Refs.~\cite{Damour:1992we,Damour:1995kt}.
In Einstein Frame, the dynamics of the scalar field arises from its coupling to the matter fields $\Psi$, and the field equations can be found in Ref.~\cite{Damour:1992we} where the parameter 
\begin{equation}
\alpha(\varphi)=\frac{\partial \ln \mathcal{A}}{\partial \varphi} \ ,
 \end{equation}
measures the coupling between the matter and the scalar field. 
The scalar field is non-minimally coupled to the metric in Jordan Frame (physical frame) 
\begin{equation}
\tilde{g}_{\mu\nu}={\mathcal{A}(\varphi)}^2 g_{\mu\nu} \ ,
\end{equation}
where $\tilde{g}_{{\mu\nu}}$ is the metric in Jordan frame. 

We follow the approach suggested by~\cite{1975ApJ...196L..59E} to ``skeletonize'' the compact, self-gravitating objects in ST theories as point particles, i.e.  the total mass of each body is dependent on the local value of the scalar field.
The skeletonized matter action with the scalar field dependent mass $\tilde{m}_I(\varphi)$  is then given by
\begin{equation}
S_{m}=-\sum_{J=A,B}\int \sqrt{-\tilde{g}_{\mu \nu}\frac{dx^{\mu}}{d\lambda}\frac{d x^{\nu}}{d\lambda}} \tilde{m}_{J}(\varphi)~,
\end{equation}
where $\lambda$ is the affine parameter.
Since $ \tilde{g}_{\mu\nu}={\mathcal{A}(\varphi)}^2 g_{\mu\nu}$, the Einstein-frame mass is defined as
\begin{equation}
m(\varphi)=\mathcal{A}(\varphi)\tilde{m}(\varphi) \ .
\end{equation}

In Paper I, we first derive the ordinary Hamiltonian (dependent only on the positions and momenta) using the contact transformation at 3PN order starting from the Lagrangian of Ref.~\cite{Bernard:2018ivi} only for the local-in-time part of the dynamics. The Jordan-Frame parameters of Ref.~\cite{Bernard:2018ivi} that encompass the scalar field effect are converted to the dimensionless Einstein-Frame parameters (see, Table I). 
The mass function $m(\varphi)$ is used to define these dimensionless body-dependent parameters following Refs.~\cite{Damour:1992we,Damour:1995kt,Julie:2017pkb} \ i.e.
\begin{align}
\alpha_I&=\frac{d\ln m(\varphi)_I}{d\varphi},\\
\beta_I&=\frac{d\alpha_I}{d\varphi},\\
\beta'_I&=\frac{d\beta_I}{d\varphi},\\
\beta''_I&=\frac{d\beta'_I}{d\varphi}~.
\end{align} 
Here, we follow the notations of Paper I for the binary parameters and use the same notation as \cite{Bernard:2018hta,Bernard:2018ivi} to denote weak-field and strong-field parameters.

Finally, we then determine the ST corrections to the EOB metric potential $(A, B, Q_e)$ at 3PN order for the local in time (instantaneous) part of the dynamics by mapping the EOB Hamiltonian in DJS gauge \cite{Damour:2000we}
\begin{align}
\label{heff-start}
\hat{H}_{\text{eff}}=\frac{H_{\text{eff}}}{\mu}=\sqrt{A(\hat{r})\left(1+\frac{\hat{p}_r^2}{B(\hat{r})}+\frac{\hat{p}_{\phi}^2}{\hat{r}^2}+q_3\frac{\hat{p}_r^4}{\hat{r}^2}\right)}~,
\end{align}
where $\hat{p}_r, \hat{p_{\phi}}$ are the dimensionless radial and angular momenta, and $\hat{r}(=r/(G_{AB}M)$ is the dimensionless radial separation, to the ordinary two-body Hamiltonian (hereafter the superscript \textit{hat} is used to denote the dimensionless variables).

The three EOB potentials at 3PN are
\begin{align}
A(\hat{r})&=1-\frac{2}{\hat{r}}+\frac{a_2}{\hat{r}^2}+\frac{a_3}{\hat{r}^3}+\frac{a_4}{\hat{r}^4}~,\\
B(r)&=1+\frac{b_1}{\hat{r}}+\frac{b_2}{\hat{r}^2}+\frac{b_3}{\hat{r}^3}~,\\
Q_e(\hat{r})&= q_3\frac{\hat{p}_r^4}{\hat{r}^2 }~.
\end{align}
The GR and ST corrections in coefficients ($a_i$, $b_i$) are separated as
\begin{align}
a_{i}&=a_{i}^{\rm GR}+\delta a_{i}^{\rm ST}~,\\
b_{i}&=b_{i}^{\rm GR}+\delta b_{i}^{\rm ST},\\
q_{3}&=q_3^{\rm GR}+\delta q_3^{\rm ST}~.
\end{align}

Since there are also nonlocal-in-time and tidal contributions at 3PN order in ST theory, all the 3PN ST coefficients can thus be decomposed as Eq.~(5.23) of Paper I.
The complete expressions of local-in-time ST corrections at 3PN can be found in Eqs.(5.14)-(5.16) of Paper I. 

In Paper I, we also derive the nonlocal-in-time (tail) and tidal corrections only for the circular orbits using the gauge invariant energy for circular orbits given in Ref.~\cite{Bernard:2018ivi, Bernard:2019yfz}. The complete expression for these coefficients can be found in Eqs. (5.25)-(5.27) of Paper I.

\section{Tail contribution to the 3PN dynamics}
\label{Sec-OrdinaryHamiltonian}
The nonlocal-in-time two-body 3PN Lagrangian for massless ST theory obtained in Ref.~\cite{Bernard:2018ivi} is in harmonic coordinates, i.e. it depends (linearly) on the acceleration of the two bodies. In this section, we will use two different methods to derive the \textit{Noetherian} conserved energy for the tail contributions. First,  we will remove the acceleration dependence from the Lagrangian (hence, the Hamiltonian) and stay within the non-order-reduced nonlocal framework (as done in Refs.~\cite{Bernard:2015njp,Bernard:2016wrg} for GR). Second, we will derive the order-reduced, local Hamiltonian using the action-angle variables (see, Ref.~\cite{Damour:2015isa} for GR).
\subsection{Non-order-reduced Ordinary Hamiltonian}
In Paper I, we derived the ordinary (dependent only on positions and momenta) Hamiltonian for local-in-time contribution using contact transformation (see, Appendix A of Paper I for the contact transformation).  Now,  concerning the nonlocal-in-time part we need to find the nonlocal shift that removes the acceleration dependence from the tail part of the Lagrangian of Ref.~\cite{Bernard:2018ivi} (see, Refs.~\cite{Bernard:2015njp,Bernard:2016wrg} for GR). Corresponding to this ordinary Lagrangian, we can then derive the ordinary Hamiltonian.

The tail part of the Lagrangian  at 3PN order reads \cite{Bernard:2018ivi},
\begin{align}
\label{tail-Bernard}
L^{\rm tail}&=\frac{2 G^2 M}{3 c^6}(3+2\omega_0)~\mathrm{Pf}_{2r_{AB}/c}\int_{-\infty}^{\infty}\frac{d\tau}{\abs{\tau}}I_{s,i}^{(2)}(t)I_{s,i}^{(2)}(t+\tau),
\end{align} 
where $\mathrm{Pf}$ is the Hadamard partie finie function, Hadamard scale $r_{AB}(=r)$ is the relative separation of two bodies, and $I_{s,i}^{(2)}$ is the second time derivative of the dipole moment. Here, we find the shift that transforms this Lagrangian into the same expression but with the derivatives of the dipole moment evaluated using the Newtonian equations of motion. In the centre of mass (COM) frame in notations of Ref.~\cite{Bernard:2018hta,Bernard:2018ivi} it is, 
\begin{align}
\label{ord-dipole}
\acute{I}_{s,i}^{(2)}=\frac{2 M\nu(s_A-s_B)}{\phi_0(3+2w_0)} \left(-\frac{G_{AB} M}{r^2}n_{AB}^i\right)~,
\end{align}
where $s_A$, $s_B$ are the sensitivity of two bodies.

As the nonlocal contribution starts at 3PN order, the ordinary Lagrangian is
\begin{align}
L_{\text{ord}}^{\text{tail}}=L^{\text{tail}}+\sum_{J=A,B}m_J\left(-a_J^i-\sum_{J\neq K} \frac{G_{AB}~m_K}{r^2} n_{JK}^i\right)\xi_{J,i}~,
\end{align}
where $L_{\text{ord}}^{\rm tail}$ is given by the same expression as Eq.~\eqref{tail-Bernard} but with second time derivative of the dipole moment replaced by its on-shell  value given in Eq.~\eqref{ord-dipole}, and the nonlocal shift, $\xi_{J,j}$,
\begin{align}
\xi_{J,j}=\frac{1}{m_J} \frac{2 G^2 M}{3c^6 }&(3+2w_0)\left[-\frac{m_J(1-2s_J)}{\phi_0(3+2w_0)}\right] \delta_j^i \nonumber \\
&\mathrm{Pf}_{2r/c} \int_{-\infty}^{\infty} \frac{d\tau}{\abs{\tau}}\acute{I}_{s,i}^{(2)}(t+\tau)~.
\end{align}

The ordinary Hamiltonian is then derived using the ordinary Legendre transformation, $H_{\text{ord}} = \sum_A p_A v_A - L_{\text{ord}}$ which reads
$H_{\rm ord}=H_{\rm ord}^{\text{loc}}+H_{\rm ord}^{\text{tail}}$, where the local contribution $H_{\rm ord}^{\text{loc}}$ is derived in Paper I (see,  Appendix C) and the tail contribution is 
\begin{align}
\label{nonlocal-Ham}
H_{\rm ord}^{\text{tail}}&=-\frac{2 G^2 M}{3c^6}(3+2w_0) ~\mathrm{Pf}_{2 r/c}\int_{-\infty}^{\infty}\frac{d\tau}{\abs{\tau}}\acute{I}_{s,i}^{(2)}(t)\acute{I}^{(2)}_{s,i}(t+\tau).
\end{align}
The tail part of the Hamiltonian is just opposite to tail part of Lagrangian.

As shown in Ref.~\cite{Damour:2016abl, Bernard:2016wrg} for the non-order-reduced, nonlocal framework the \textit{Noetherian} conserved energy ($E_{\rm cons}$) is not given by the Hamiltonian but is given by, $E_{\rm cons}=H_{\rm ord}^{\text{tail}}+\delta H$. This additional term $\delta H$ consists of purely a constant term (DC type) and time oscillating term with zero average value (AC type) and is same as given in Eq. (4.10) of Ref.~\cite{Bernard:2018ivi}.
\subsection{Order-reduced Ordinary Hamiltonian}
\label{Action-angleH}
The second method to derive the conserved energy for tail part is to work in the order-reduced, local framework as given in Ref.~\cite{Damour:2015isa,Damour:2016abl} for GR. 

The tail part of the Hamiltonian in ST theory is,
\begin{align}
\label{tail-BernardH}
H^{\rm tail}=-\frac{2 G^2 M}{3 c^6}(3+2\omega_0)&\left[\mathrm{Pf}_{2r/c}\int_{-\infty}^{\infty}\frac{d\tau}{\abs{\tau}}I_{s,i}^{(2)}(t)I_{s,i}^{(2)}(t+\tau)\right.\nonumber \\
&\left.-2\ln\left(\frac{\hat{r}}{a}\right)I_{s,i}^{(2)}(t)^2\right] .
\end{align} 
As mentioned in Ref.~\cite{Bernard:2016wrg}, in the action-angle form there should be an additional term (second term in Eq.~\eqref{tail-BernardH}) which is local and accounts for dependence of Hadamard Partie finie function on the radial separation ($r$) at time $t$ i.e., $\hat{r} = a(1-e\cos(u))$ in action-angle variables.

The basic methodology we use to order-reduce the nonlocal dynamics of the above form is based on Refs.~\cite{Damour:2015isa,Damour:2016abl} for GR, and consists of four main steps: (i) Re-express the Hamiltonian in terms of action angle variables, (ii)``order-reduce" the nonolocal dependence on action angle variable, (iii) expand it in powers of eccentricity, and (iv) eliminate the periodic terms in order-reduced Hamiltonian by a canonical transformation. All of these steps lead to the order-reduced ordinary local Hamiltonian for the tail part in terms of action-angle variables.

Let us consider the expression of nonlocal-in-time piece of Eq.~\eqref{tail-BernardH}, \ i.e.
\begin{align}
\label{nonloc-expr}
\mathcal{K}(t,\tau)=\ddot{I}_{s,i}(t)\ddot{I}_{s,i}(t+\tau)~.
\end{align}
To order reduce the nonlocal piece, we use the equations of motion to express the phase-space variables at shifted time $t+\tau$ in terms of the phase-space variables at time $t$. As the zeroth order equations are Newtonian equations, it will be convenient to use the action-angle form of the Newtonian equations of motion,
\begin{align}
\label{actioneq}
\frac{\partial l}{\partial \hat{t}}&=\frac{\partial H_0}{\partial \mathcal{L}}=\frac{1}{\mathcal{L}^3}=\Omega(\mathcal{L})~,~~\frac{\partial \mathcal{L}}{\partial \hat{t}}=\frac{\partial H_0}{\partial l}=0~,\nonumber \\
\frac{\partial \mathcal{G}}{\partial \hat{t}}&=\frac{\partial H_0}{\partial g}=0~,\hspace{1.7cm}\frac{\partial g}{\partial \hat{t}}=\frac{\partial H_0}{\partial \mathcal{G}}=0~,
\end{align}
where $\hat{t}={t}/({G_{AB}M})$ is the dimensionless time variable, $(\mathcal{L},l,\mathcal{G},g)$ are the action-angle variables. The zeroth-order (Newtonian) Hamiltonian in action-angle variable is  $H_0=-1/(2\mathcal{L}^2)$.

Here, the variable $\mathcal{L}$ is conjugate to the ``mean anamoly" $l$ and $\mathcal{G}$ is conjugate to argument of periastron $g$. In terms of the Keplerian variables, semi-major axis $a$, and eccentricity $e$, these are
\begin{align}
\mathcal{L}=\sqrt{a},\hspace{0.4cm}\mathcal{G}=\sqrt{a(1-e^2)}~.
\end{align}

From Eq.~\eqref{actioneq}, the variables $\mathcal{L}$, $\mathcal{G}$ and $g$ are independent of time, and $l$ varies linearly with time, hence it will be sufficient to use
\begin{align}
l(t+\tau)=l(t)+\Omega~ \hat{\tau}~,
\end{align}
where $\hat{\tau}=\tau/(G_{AB}M)$. The order-reduced non-local in time expression of Eq.~\eqref{nonloc-expr} becomes
\begin{align}
\mathcal{K}(t,\tau)&=\left(\frac{1}{G_{AB}M}\right)^4\mathcal{K}(\hat{t},\hat{\tau})\nonumber\\
&=\left(\frac{\Omega}{G_{AB}M}\right)^4\frac{d^2}{dl^2} {I}_{s,i}(l)\frac{d^2}{dl^2}{I}_{s,i}(l+\Omega \hat{\tau})~.
\end{align}

Using the Fourier decomposition of dipole moment given in Eq.~\eqref{Fouier-exp}, we find the structure of nonlocal-in-time expression $\mathcal{K}(t,\hat{\tau})$ and hence the Hamiltonian. As shown in \cite{Damour:2015isa} for GR, all the periodically varying terms can be eliminated by a suitable canonical transformation. Hence, the order-reduced Hamiltonian can be further simplified by replacing $H^{\rm tail}$ with its $l$-average value
\begin{align}
\label{order-red_ham}
\bar{H}_{\text{tail}}=\int_0^{2 \pi} dl H^{\rm tail}~.
\end{align}
%

Using the result
\begin{align}
\mathrm{Pf}_T\int_0^{\infty}\frac{dv}{v}\cos(\omega v)=-(\gamma_{\rm E}+\ln(\omega~T))~\hspace{1.4cm}\forall~(\omega>0)~,
\end{align}
where $\gamma_{\rm E}$ is the Euler's constant, and inserting the expression of $r$ from Eq.~\eqref{action-angleformr}, the Hamiltonian, Eq.~\eqref{order-red_ham}, reads
\begin{align}
\label{nonloc-f}
\bar{H}_{\text{tail}}=\frac{8 G^2 M}{3}\left(\frac{\Omega}{G_{AB}M}\right)^4&(3+2w_0)\sum_{p=1}^{\infty}p^4\abs{I_{s,i}(p)}^2 \nonumber \\
&\ln\left(\mathrm{e}^{\rm \gamma_{\rm E}}\frac{2p~a~\Omega }{c}\right).
\end{align}

Now, inserting the Fourier-Bessel expansion of scalar dipole moment from Eqs.~\eqref{BesselIx}-\eqref{BesselIy} (see, Appendix \ref{Fourier-Bderived} for derivation) in Eq.~\eqref{nonloc-f}, the real two-body nonlocal-in-time Hamiltonian in order-reduced, local framework is (in notations of Paper I)
\begin{widetext}
\begin{align}
\hat{\bar{H}}_{\text{tail}}\equiv\frac{\bar{H}_{\text{tail}}}{\mu}=\frac{2\nu}{3a^4}\left(2 \delta_+ +\frac{\bar{\gamma}_{AB}(\bar{\gamma}_{AB}+2)}{2}\right)\sum_{p=1}^{\infty}\frac{p^2}{e^2}&\left\{4e^2J^2_{p-1}(pe)+(8-4e^2) J^2_p(pe)-8eJ_{p-1}(pe)J_p(pe)\right\}\nonumber\\
& \left[\gamma_{\rm E}+\ln\left(\frac{2p~a^{-1/2}}{c}\right)\right]~.
\end{align}
\end{widetext}
Expanding the result in powers of eccentricity, the Hamiltonian as an expansion in eccentricity upto order of $e^4$ reads
\begin{widetext}
\begin{align}
\hat{\bar{H}}_{\text{tail}}&=\frac{2\nu}{3a^4}\left(2 \delta_+ +\frac{\bar{\gamma}_{AB}(\bar{\gamma}_{AB}+2)}{2}\right)\left\{2\ln(2)-\ln(a)+2\gamma_{\rm E}+e^2\left(14\ln(2)+6\gamma_{\rm E}-3\ln(a)\right)\right.\nonumber\\
&\left.+e^4\left(\frac{45}{4}\gamma_{\rm E}-\frac{3}{4}\ln(2)+\frac{729}{32}\ln(3)-\frac{45}{8}\ln(a)\right)+\mathcal{O}(e^6)\right\}~.
\end{align}
\end{widetext}
\section{Scalar Tensor corrections to Effective One Body at 3PN:Tail}
\label{Sec-EOB}
In this section, we will derive the complete tail corrections to the EOB metric 
potentials $(A,B,Q_e)$ for ST theories at 3PN order.   

Similar to the decomposition of complete 3PN coefficient $\delta a_4^{\text{ST}}$ 
in Eq. (5.23) of Paper I, we decompose the complete 
3PN ST coefficients $\delta b_3^{\text{ST}},\delta q_3^{\text{ST}}$ as
\begin{align}
\label{b-decompose}
\delta b_3^{ST}&=\delta b_{3,\rm loc}^{\rm ST}+\delta b_{3,\rm nonloc}^{\rm ST}+\delta b_{3,\rm tidal}^{\rm ST}~,\\
\label{q-decompose}
\delta q_3^{ST}&=\delta q_{3,\rm loc}^{\rm ST}+\delta q_{3,\rm nonloc}^{\rm ST}+\delta q_{3,\rm tidal}^{\rm ST}~,
\end{align}
where the local contributions $(\delta b_{3,\rm loc}^{\rm ST},\delta q_{3,\rm loc}^{\rm ST})$ are derived in Paper I (see, Eqs.  (5.14)-(5.15)),  
$(\delta b_{3,\rm nonloc}^{\rm ST},\delta q_{3,\rm nonloc}^{\rm ST})$ are the nonlocal contributions, 
and $(\delta b_{3,\rm tidal}^{\rm ST},\delta q_{3,\rm tidal}^{\rm ST})$ are the tidal contributions. 
The nonlocal contributions can be further decomposed similar to Eq. (5.24) of Paper as
\begin{align}
\label{tail-decompose} 
\delta a_{4,\rm nonloc}^{\rm ST}&=\delta a_{4,\rm nonloc,0}^{\rm ST}+\delta a_{4,\rm nonloc,\rm log}^{\rm ST}\ln (\hat{r})~,\\
\label{b-tail}
\delta b_{3,\rm nonloc}^{\rm ST}&=\delta b_{3,\rm nonloc,0}^{\rm ST}+\delta b_{3,\rm nonloc,\rm log}^{\rm ST}\ln(\hat{ r})~,\\
\label{q-tail}
\delta q_{3,\rm nonloc}^{\rm ST}&=\delta q_{3,\rm nonloc,0}^{\rm ST}+\delta q_{3,\rm nonloc,\rm log}^{\rm ST}\ln(\hat{r})~.
\end{align}

Inserting the split of the EOB functions ($A$, $B$, $q_3$) using 
Eqs.~\eqref{b-decompose}-\eqref{q-decompose} and Eq. (5.23) of Paper I
in the effective Hamiltonian of Eq. \eqref{heff-start}, 
and then after expanding the right-side into a Taylor series of $1/c^2$, we obtain
\begin{align}
\hat{H}_{\rm eff}=\hat{H}^{\rm loc}_{\rm eff}+\hat{H}^{\rm nonloc}_{\rm eff}~,
\end{align}
where $\hat{H}^{\rm loc}_{\rm eff}$ is computed only by the local contributions 
$(\delta a_{4,\rm loc}^{\rm ST},  \delta b_{3,\rm loc}^{\rm ST},\delta q_{3,\rm loc}^{\rm ST})$ and 
$\hat{H}^{\rm nonloc}_{\rm eff}$ is the nonlocal contribution of Hamiltonian computed by $(\delta a_{4,\rm nonloc}^{\rm ST},  \delta b_{3,\rm nonloc}^{\rm ST},\delta q_{3,\rm nonloc}^{\rm ST})$. 
The nonlocal contribution $\hat{H}^{\rm nonloc}_{\rm eff}$ reads
\begin{align}
\label{heff}
\hat{H}^{\rm nonloc}_{\rm eff}=\frac{1}{2}\left(\delta a_{4,\rm nonloc}^{\rm ST}\frac{1}{\hat{r}^4}-\delta b_{3,\rm nonloc}^{\rm ST}\frac{\hat{p}_r^2}{\hat{r}^3}+\delta q_{3,\rm nonloc}^{\rm ST}\frac{\hat{p}_r^4}{\hat{r}^2}\right)~.
\end{align}

To map the real two-body dynamics to EOB, we express the nonlocal effective Hamiltonian, $\hat{H}^{\rm nonloc}_{\rm eff}$, 
in action-angle variables $\mathcal{L}$, $l$, $\mathcal{G}$, and $g$ (hence the Keplerian variables $a$ and $e$) 
and compute its $l$-averaged value,
\begin{align}
\hat{\bar{H}}^{\rm nonloc}_{\rm eff}=\frac{1}{2\pi}\int_0^{2\pi}d l \hat{H}^{\rm nonloc}_{\rm eff}~.
\end{align}

The explicit expression of $\hat{H}^{\rm nonloc}_{\rm eff}$ depends on $l$-average monomials involving powers of 
$1/\hat{r}$ and $\hat{p}_r$ (and also $\ln(\hat{r})$ from Eqs.~\eqref{tail-decompose}, \eqref{b-tail}, and \eqref{q-tail}).  
These computations can be performed by expanding Eq.~\eqref{heff} in terms of eccentricity upto $e^5$ 
using the Newtonian equations of motion in action-angle form 
recalled in Sec.~\ref{Action-angleH}. 
The $l$-averaged value we obtain is
\begin{widetext}
\begin{align}
\label{last}
\hat{\bar{H}}^{\rm II}_{\rm eff}&=\frac{1}{2a^4}\left\{\delta a_{4,\rm nonloc,0}^{\rm ST}+\delta a_{4,\rm nonloc,\rm log}^{\rm ST}\ln(a)\right.\nonumber\\
&\left.+\left(3\delta a_{4,\rm nonloc,0}^{\rm ST}-\frac{7}{4}\delta a_{4,\rm nonloc,\rm log}^{\rm ST}-\frac{1}{2}\delta b_{3,\rm nonloc,0}^{\rm ST}+3\delta a_{4,\rm nonloc,\rm log}^{\rm ST}\ln(a)-\frac{1}{2}\delta b_{3,\rm nonloc,\rm log}^{\rm ST}\ln(a)\right)e^2\right.\nonumber\\
&\left.+\left( \frac{45}{8}\left[\delta a_{4,\rm nonloc,0}^{\rm ST}+\delta a_{4,\rm nonloc,\rm log}^{\rm ST}\ln(a)\right]-\frac{5}{4}\left[\delta b_{3,\rm nonloc,0}^{\rm ST}+\delta b_{3,\rm nonloc,\rm log}^{\rm ST}\ln(a)\right]+\frac{3}{8}\left[\delta q_{3,\rm nonloc,0}^{\rm ST}+\delta q_{3,\rm nonloc,\rm log}^{\rm ST}\ln(a)\right]\right.\right.\nonumber\\
&\left.\left.-\frac{171}{32}\delta a_{4,\rm nonloc,\rm log}^{\rm ST}+\frac{9}{16}\delta b_{3,\rm nonloc,\rm log}^{\rm ST}\right)e^4+\mathcal{O}(e^6)\right\}~.
\end{align}
\end{widetext}

The final step is then to map the real two-body dynamics to EOB metric by the \textit{nontrivial} map,
\begin{align}
\hat{H}_{\rm real}=\frac{H_{\rm real}}{\mu}=\frac{1}{\nu}\sqrt{1+2\nu(\hat{H}_{\rm eff}-1)}~,
\end{align}
between the EOB Hamiltonian $(\hat{H}_{\rm eff})$ and real two-body Hamiltonian $(\hat{H}_{\rm real})$.  
The quadratic map relating the two Hamiltonians is proven at \textit{all} PN orders in GR and ST within the Post-Minkowskian scheme in Ref.~\cite{Damour:2016gwp}.  However, it can be seen that only for the nonlocal contributions at 3PN order, 
the map relating the two nonlocal Hamiltonians is 
\begin{align}
\hat{\bar{H}}^{\rm nonloc}_{\rm eff}=\hat{\bar{H}}^{\rm II}_{\rm real, nonloc}~.
\end{align}

The \textit{unique} nonlocal ST contributions at 3PN from this matching are
\begin{widetext}
\begin{align}
\label{circ-0}
\delta a^{\rm ST}_{4,\rm nonloc,0}&=\frac{4}{3}\nu\left[2\delta_++\frac{\bar{\gamma}_{AB}(\bar{\gamma}_{AB}+2)}{2}\right](2\ln 2+2\gamma_{\rm E}),\\
\label{circ-log}
\delta a^{\rm ST}_{4,\rm nonloc,\rm log}&=-\frac{4}{3}\nu\left[2\delta_++\frac{\bar{\gamma}_{AB}(\bar{\gamma}_{AB}+2)}{2}\right],\\
\delta b_{3,\rm nonloc,0}^{\rm ST}&=\frac{4}{3}\nu\left[2\delta_++\frac{\bar{\gamma}_{AB}(\bar{\gamma}_{AB}+2)}{2}\right]\left(\frac{21}{2}-16\ln 2\right),\\
\delta b_{3,\rm nonloc,\rm log}^{\rm ST}&=0,\\
\delta q_{3,\rm nonloc,0}^{\rm ST}&=\frac{4}{3}\nu\left[2\delta_++\frac{\bar{\gamma}_{AB}(\bar{\gamma}_{AB}+2)}{2}\right]\left(-\frac{31}{4}-\frac{256}{3}\ln 2+\frac{243}{4}\ln 3\right),\\
\delta q_{3,\rm nonloc,\rm log}^{\rm ST}&=0~.
\end{align}
\end{widetext}

The ST tensor correction $\delta a^{\rm ST}_{4,\rm nonloc}$ for the circular orbit case, Eqs.~\eqref{circ-0}-\eqref{circ-log}, matches with the results obtained in Paper I (see, Eqs.(5.25)-(5.26)) except a negative sign in Eq.~\eqref{circ-log}.  The negative sign is due to the difference in the definition of $\delta a^{\rm ST}_{4,\rm nonloc}$ in Eq.~\eqref{tail-decompose} used in this work with the Eq.~(5.24) of Paper I. 
\section{Conclusions}
In Paper I, building upon the results of \cite{Bernard:2018ivi} for massless scalar-tensor theory, we determined the EOB coefficients at 3PN order though restricting ourselves to local-in-time part of the dynamics and nonlocal-in-time and tail contributions only for the circular case. In the present paper, we derived the complete nonlocal-in-time EOB coefficients starting from the nonlocal-in-time Lagrangian of Ref.~\cite{Bernard:2018ivi}. First, we derived the two-body  \textit{conserved} ordinary Hamiltonian (dependent only on positions and momenta) for nonlocal-in-time part by two methods: (i) non-order-reduced nonlocal Hamiltonian using nonlocal phase shift (see, Ref.~\cite{Bernard:2015njp,Bernard:2016wrg} for GR), and (ii) order-reduction of nonlocal dynamics to local ordinary action-angle Hamiltonian \cite{Damour:2015isa}. We then expressed the effective Hamiltonian in Delaunay variables to recast the order-reduced ordinary action-angle Hamiltonian into equivalent, 3PN-accurate, nonlocal part of EOB potentials $(A, B, Q_e)$, see Eqs. (4.12)-(4.17). 

By combining the results of Paper I and the present work, we could transcribe the two-body Hamiltonian into equivalent 3PN-accurate EOB potentials $(A, B, Q_e)$ for both local-in-time and nonlocal-in-time part of dynamics.
\\
\\
\textbf{Note:} During the preparation of the final manuscript of this work, the author became aware of the independent effort which recently arrived on arXiv \cite{Julie:2022qux}.
\begin{acknowledgements} 
The author is grateful to P. Rettegno, M. Agathos and A. Nagar for useful discussions and suggestions during the preparation of this work. The author is jointly funded by the University of Cambridge Trust, Department of Applied Mathematics and Theoretical Physics (DAMTP), and Centre for Doctoral Training, University of Cambridge. 
\end{acknowledgements}
\appendix
\section{Fourier Coefficients of dipole moment in ST theory}
\label{Fourier-Bderived}
In this appendix, we will determine the explicit expressions of \textit{Newtonian} 
dipole moment in ST theory using the known Fourier decomposition 
of the Keplerian motion (see, Refs.~\cite{Peters:1963ux, Arun:2007rg} for GR). 

The dipole moment, $I_{s,i}(t)$, in COM frame is 
\begin{align}
I_{s,i}(t)=\frac{2 M\nu(s_A-s_B)}{\phi_0(3+2w_0)}~ \mathrm{x}_i~,
\end{align}
where $\mathrm{x}_i=(Z_A-Z_B)_i$ is the relative separation vector and $Z_{A,B}$ indicate the positions of the two bodies.

Since the motion is planar, we can choose the coordinate system $(x,y,z)$ such that it coincides with the $xy$-plane. Using the polar coordinates $(\hat{r},\phi_a)$,
\begin{align}
x=\hat{r}~\cos(\phi_a),\hspace{0.2cm} y=\hat{r}~\sin(\phi_a).
\end{align}
The coordinates $(x,y)$ are the coordinates of the dimensionless relative separation, $\hat{r}=x_A-x_B$ with ${x}_J={\mathrm{x}}_J/(G_{AB}M)$ denoting the position of two bodies.

As mentioned in Ref.~\cite{Peters:1963ux,Arun:2007rg, Damour:2015isa} for GR,  for leading order contributions it is convenient to use the Delaunay (action-angle) form of the Newtonian equations of motion.  In terms of the action-angle variables $(\mathcal{L},l,\mathcal{G},g)$, the Cartesian coordinates $(x,y)$ are given by (Here, we follow the notations of \cite{CelesMech})

\begin{align}
\label{action-angleformx}
x&= x_0\cos(g)-y_0\sin(g)~,\\
\label{action-angleformy}
y&=x_0\cos(g)+y_0\sin(g)~,\\
\label{action-angleformx0}
x_0&=\hat{r} \cos(f)=a(\cos(u)-e)~,\\
\label{action-angleformy0}
y_0&=\hat{r}\sin(f)=a\sqrt{1-e^2}\sin(u)~,\\
\label{action-angleformr}
\hat{r}&=a(1-e\cos(u))~,
\end{align}
where $a$ is the semi-major axis, $e$ is the eccentricity, $f$ is the ``true anamoly" and  the ``eccenteric anamoly" $u$ in terms of Bessels functions is given by
\begin{align}
\label{action-angleu}
u=l+\sum_{n=1}^{\infty}\frac{2}{n}\mathrm{J}_n(n e) \sin(nl)~.
\end{align}
The Bessel-Fourier expansion of $\cos(u)$ and $\sin(u)$, which directly enters $x_0,y_0$ are:
\begin{align}
\label{cosuBessel}
\cos(u)&=-\frac{e}{2}+\sum_{n=1}^{\infty}\frac{1}{n}\left[\mathrm{J}_{n-1}(ne)-\mathrm{J}_{n+1}(ne)\right]\cos(nl)~,\\
\label{sinuBessel}
\sin(u)&=\sum_{n=1}^{\infty}\frac{1}{n}\left[\mathrm{J}_{n-1}(ne)+\mathrm{J}_{n+1}(ne)\right]\sin(nl)~.
\end{align}

From Eqs.~\eqref{action-angleformx}-\eqref{action-angleu}, the dipole moment $I_{s,i}$ is a periodic function of $l$ (and hence time) at the Newtonian order. Thus it can be decomposed into Fourier series
\begin{align}
\label{Fouier-exp}
I_{s,i}(l)=\sum_{p=-\infty}^{\infty}I_{i,s}(p)~\mathrm{e}^{i  pl}~,
\end{align}
with
\begin{align}
\label{Fourier-coeff}
I_{s,i}(p)=\frac{1}{2\pi}\int_{0}^{2\pi}dl ~I_{s,i} ~\mathrm{e}^{-ipl}~.
\end{align}

The Fourier coefficients of the scalar dipole moment at the Newtonian order are derived using Eq.~\eqref{Fourier-coeff} in terms of combinations of Bessel Functions. 

Inserting the expression of Cartesian coordinates in terms of action-angle variables using Eqs.~\eqref{action-angleformx}-\eqref{sinuBessel},  we find the Fourier-Bessel coefficients of the scalar dipole moment are
\begin{widetext}
\begin{align}
\label{BesselIx}
I_{s,x}(p)&= G_{AB} M\left[\frac{2 M\nu(s_A-s_B)}{\phi_0(3+2w_0)}\frac{a}{2p}\left\{\left[J_{p-1}(p e)-J_{p+1}(pe)\right]\cos(g)+i\sqrt{1-e^2}\left[J_{p-1}(pe)+J_{p+1}(pe)\right]\sin(g)\right\}\right]~,\\
\label{BesselIy}
I_{s,y}(p)&=G_{AB} M \left[ \frac{2 M\nu(s_A-s_B)}{\phi_0(3+2w_0)}\frac{a}{2p}\left\{\left[J_{p-1}(p e)-J_{p+1}(pe)\right]\sin(g)+i\sqrt{1-e^2}\left[J_{p-1}(pe)+J_{p+1}(pe)\right]\cos(g)\right\}\right]~.
\end{align}
\end{widetext}
\bibliography{local_tail.bib,refs.bib}
\end{document}